\documentstyle[prb,aps,epsf,epsfig]{revtex}

\begin{document}

\twocolumn[\hsize\textwidth\columnwidth\hsize\csname@twocolumnfalse\endcsname
\title{The puzzle of 90$^{\circ }$ reorientation in the vortex lattice of borocarbide superconductors}
\author{Anton Knigavko and Baruch Rosenstein}
\address{Department of Electrophysics, National Chiao Tung University,
Hsinchu, Taiwan 30050, Republic of China}
\maketitle

\begin{abstract}
We explain 90$^{\circ }$ reorientation in the vortex lattice of borocarbide
superconductors on the basis of a phenomenological extension of the nonlocal
London model that takes full account of the symmetry of the system. We
propose microscopic mechanisms that could generate the correction terms and
point out the important role of the superconducting gap anisotropy.
\end{abstract}


]


Abrikosov vortices in type two superconductors repel each other and
therefore tend to form two dimensional lattices when thermal fluctuations or
disorder are not strong enough to destroy lateral correlations. In isotropic
s--wave materials the lattices are triangular, however in anisotropic
materials or for ''unconventional'' d--wave or p--wave pairing interactions
less symmetric vortex lattices (VL) can form as recent experiment on high $%
T_{c}$ cuprates \cite{cuprate}, $SrRuO_{4}$\cite{rutenate} and borocarbides
showed. The quality of samples in the last kind of superconductors allows
detailed reconstruction of the phase diagram by means of small angle neutron
scattering, scanning tunnelling microscopy or Bitter decoration technique.
For $H||c$ the presence of a whole series of structural transformations of
VL was firmly established. At first, stable at high magnetic fields square
lattice becomes rhombic, or ''distorted triangular'', via a second order
phase transition \cite{squre-rhomb,stm}. Then, at lower fields, 45$^{\circ }$
reorientation of VL relative to crystal axis occurs\cite{PTI,45degree}. For $%
H||a$ a continuous lock-in phase transition was predicted\cite{PTI}. Above
this transition apex angle of elementary rhombic cell of VL does not depend
on magnetic field, but below a critical field such a dependence appears.

Theoretically the mixed state in nonmagnetic borocarbide superconductors $%
RNi_{2}B_{2}C$, $R=Y,Lu$ can be understood in the frame of the extended
London model \cite{ext-L} (in regions of the phase diagram close to $%
H_{c2}(T)$ line the extended Ginzburg--Landau model can be used \cite
{stm,ext-GL}). So far this theory always provided qualitative and even
quantitative description of phase transitions in VL and various other
properties such as magnetization behavior \cite{magnetization}, dependence
of nonlocal properties on the disorder\cite{Co}, etc. However, recently
another ''reorientation'' phase transition has been clearly observed in
neutron scattering experiment on $LuNi_{2}B_{2}C,$ which cannot be explained
by the theory despite considerable efforts. When magnetic field of $0.3T$
was applied along the $a$ axis of this tetragonal superconductor sudden 90$%
^{\circ }$ reorientation of VL has been seen \cite{90reo-neut}. At this
point a rhombic (nearly hexagonal, apex angle $\approx 60^{\circ }$)
lattice, oriented in such a way that the crystallographic axes are its
symmetry axes, gets rotated by $90^{\circ }.$ Both initial and rotated
lattices are found to coexist at the field range of about $0.1T$ wide around
the transition. Similar observations have been made in magnetic material $%
ErNi_{2}B_{2}C.$

In this Letter we explain why the extended London model in its original form
cannot generally explain even the existence of the $90^{\circ }$
reorientation transition. The reason is that it possesses a ''hidden''
spurious fourfold symmetry preventing such a transition. Then we generalize
the model to include the symmetry breaking effect and explain why the
reorientation take place. Then we search for a microscopic origin of this
effect. Using BCS type theory we find that anisotropy of the Fermi surface
is ruled out due to smallness of its contribution. It is anisotropy of the
pairing interaction that provides the required mechanism. We, therefore,
suggest that there exist a correlation between the critical field of 90$%
^{\circ }$ reorientation in VL and the value of the anisotropy of the gap.


A convenient starting point of any generalized ''London'' model \cite
{ext-L,Franz} is the linearized relation between the supercurrent $j_{i}$
and the vector potential $A_{j}$: 
\begin{equation}
(4\pi/c) j_{i}({\bf q})=-K_{ij}({\bf q})A_{j}({\bf q).}
\label{kernel-general}
\end{equation}
In the standard London limit the kernel $K_{ij}({\bf q})$ is approximated
just by its $q=0$ limit, inverse mass matrix, while in the extended London
model the quadratic terms of the expansion of the kernel near $q=0$ are also
kept\cite{ext-L}: 
\begin{equation}
K_{ij}({\bf q})=m_{ij}^{-1}/\lambda ^{2}+n_{ij,kl}q_{k}q_{l}.
\label{kernel-ext-L}
\end{equation}
The significance of the quantity $n_{ij,kl}$ is that it allows proper
account of the symmetry of any crystal system while the first term does not
guarantee this. At the same time it expresses nonlocal effects which are
inherent to the electrodynamics of superconductors and below we call its
component or their combination nonlocal parameters. From its definition, $%
n_{ij,kl}\equiv \frac{1}{2}\frac{\partial ^{2}}{\partial q_{l}\partial q_{m}}%
K_{ij}(q)|_{q=0}$ is a tensor with respect to both the first and the second
pairs of indices. However, the way $n_{ij,kl}$ transforms when the first and
the second pairs of indices are interchanged is not obvious because the
''origin'' of these indices are quite different. The first pair $(ij)$
comes, roughly speaking, from the variation of the free energy of the system
''a superconductor in weakly inhomogeneous magnetic field'' with respect to
the vector potentia while the second pair $(kl)$ comes from the expansion in
vector ${\bf q.}$ Below we show that in general no symmetry $%
n_{ij,kl}=n_{kl,ij}$ is expected.

The original derivation Eq. (\ref{kernel-ext-L}) from BCS theory in
quasiclassical Eilenberger formulation\cite{ext-L} produced a fully
symmetric rank four tensor: $n_{ij,kl}\sim \langle
v_{i}v_{j}v_{k}v_{l}\rangle $ with $v_{i}$ being components of velocity of
electrons at the Fermi surface. In this calculation independence of the gap
function on the orientation was assumed. Let us consider vortex lattice
problem with this result. Specializing to tetragonal borocarbides, the
number of independent component of tensor $n_{ij,kl}$ is four: $n_{aaaa,}$ $%
n_{aabb},$ $n_{aacc}$ and $n_{cccc}.$ In the case of external magnetic field
oriented along $a$ axis the free energy of VL, which is the relevant
thermodynamic potential for a thin plate sample in perpendicular external
field, reads 
\begin{eqnarray}
F &=&\left( B^{2}/8\pi \right) \sum \left[ 1+D(g_{x},g_{y})\right] ^{-1},
\label{Ext-Lond-Energy} \\
D &=&\lambda ^{2}(m_{a}g_{x}^{2}+m_{c}g_{y}^{2})+\lambda ^{4}\left[
n(m_{a}g_{x}^{2}+m_{c}g_{y}^{2})^{2}+dg_{x}^{2}g_{y}^{2}\right] .  \nonumber
\end{eqnarray}
Here $B$ is magnetic induction and the summation runs over all vectors ${\bf %
g}$ of the reciprocal VL. The nonlocal parameters appearing in this equation
have the form $n=n_{aacc}$ and $%
d=n_{cccc}m_{c}^{2}+n_{aaaa}m_{a}^{2}-6n_{aacc}m_{a}m_{c}.$ The free energy
of Eq. (\ref{Ext-Lond-Energy}) has been extensively studied numerically
first minimizing it on the class of rhombic lattices with symmetry axes
coinciding with the crystallographic axes \cite{PTI} and more recently by us
for arbitrary lattices with one flux per unit cell. Despite the fact that
great variety of \ vortex lattice transformation were identified, no a $%
90^{\circ }$ reorientation has been ever seen. The reason is quite simple:
the considered free energy is actually effectively fourfold symmetric. After
rescaling the reciprocal lattice vectors 
\begin{equation}
g_{x}\rightarrow \tilde{g}_{x}\equiv g_{x}/\sqrt{m_{a}},\;g_{y}\rightarrow 
\tilde{g}_{y}\equiv g_{y}/\sqrt{m_{c}}  \label{scale-transf}
\end{equation}
the sum in Eq.(\ref{Ext-Lond-Energy}) becomes fourfold symmetric explicitly.
Based on this observation one concludes that energies of the lattices
participating in the $90^{\circ }$ reorientation are equal exactly.
Therefore no phase transition between them is possible in the framework of
the extended London model of Eq.(\ref{Ext-Lond-Energy}) and further
corrections are necessary to account for this transition.

There might be a slight possibility that the observed $90^{\circ }$
reorientation presents the lock-in transition described in the beginning of
this paper. For this to happen the rescaled square VL should looks almost
hexagonal and, correspondingly, a particular value of masses asymmetry $%
m_{a}/m_{c}=\left[ \cos \left( 60^{\circ }\right) /\cos \left( 45^{\circ
}\right) \right] ^{2}=1/2$ is required. This is very different from the
figures quoted in literature \cite{PTI}: $m_{a}/m_{c}=0.9/1.22$ $=0.74$.
More importantly, according to this scenario one should see two degenerate
lattices at small fields below the transition and only a single lattice at
high fields above the transition which experimentally is clearly not the
case.


To explain $90^{\circ }$ reorientation we proceed by correcting the model of
Eq. (\ref{Ext-Lond-Energy}). On general symmetry grounds for $H||a$ one can
expect more terms in the expression for $D$ which describes vortex-vortex
interactions. Given two fold symmetry of the present case we write down for $%
D$ the expansion in Fourier series up to fourth harmonics, perform rescaling
defined by Eq. (\ref{scale-transf}) and obtain 
\begin{equation}
D_{eff}=D_{0}(\tilde{g})+D_{4}(\tilde{g})\cos (4\varphi )+D_{2}(\tilde{g}%
)\cos (2\varphi ),  \label{Fourier-exp}
\end{equation}
where $\varphi $ is the polar angle in the rescaled $b-c$ plane. The
quantity $D$ from Eq. (\ref{Ext-Lond-Energy}) produces only fourfold
invariant terms: 
\begin{eqnarray}
D_{0}(\tilde{g}) &=&\lambda ^{2}\tilde{g}^{2}+\left( n+d/8m_{a}m_{c}\right)
\lambda ^{4}\tilde{g}^{4}, \\
D_{4}(\tilde{g}) &=&-\left( d/8m_{a}m_{c}\right) \lambda ^{4}\tilde{g}^{4}.
\end{eqnarray}
The new term $D_{2}(\tilde{g})$ expresses the effective fourfold symmetry
breaking. Experimentally, it should be small as indicated by recent success
in qualitative understanding the angle dependence of magnetization of $%
LuNi_{2}B_{2}C$ \cite{magnetization} with field lying in the $a-b$ plane on
the basis of the theory without $D_{2}$ term. Accordingly, we can treat it
perturbatively:$F=F^{(0)}+F^{^{(pert)}}$ with 
\begin{eqnarray}
F^{(0)} &=&\left( B^{2}/8\pi \right) \sum \left[ 1+D_{0}+D_{4}\cos (4\varphi
)\right] ^{-1},  \label{Eff-ener-zero} \\
F^{(pert)} &=&-\left( B^{2}/8\pi \right) \sum \frac{D_{2}\cos (2\varphi )}{%
\left[ 1+D_{0}+D_{4}\cos (4\varphi )\right] ^{2}}  \label{Eff-ener-pert}
\end{eqnarray}
where the summation is over ${\bf \tilde{g}}$ (see Eq. (\ref{scale-transf}%
)). The original degeneracy of two VL rotated by $90^{\circ }$ with respect
to each other is split now. To explain the $90^{\circ }$ reorientation the
sign of the perturbation should change at certain field $B_{reo}$. Magnetic
field influences the sum via constraint that area of the unit cell carries
one fluxon. Roughly speaking $D_{2}(g)$ should change sign when $\tilde{g}%
\approx \sqrt{B_{reo}/\Phi _{0}}.$ The simplest way to implement this idea
is to write for $D_{2}(g)$ two lowest order terms in $\tilde{g}:$%
\begin{eqnarray}
D_{2}=w_{4}\tilde{g}^{4}+w_{6}\tilde{g}^{6}  \label{D2}
\end{eqnarray}
Quadratic term is not present since we have already rescaled it out in
derivation of Eq. (\ref{Fourier-exp}). In principle the coefficient $w_{6}$
can be derived from BCS similarly to $n_{ij,kl}$ tensor within the framework
of original extended London model \cite{ext-L}. Then it is proportional to
the Fermi surface average of six components of Fermi velocity. To obtain $%
w_{4},$ however, the result $n_{ij,kl}\sim \langle
v_{i}v_{j}v_{k}v_{l}\rangle $ of Ref. \onlinecite{ext-L} is not sufficient.
Indeed, using general expression for $n_{ij,kl}$ and repeating derivation of
Eq. (\ref{Ext-Lond-Energy}) from Eq. (\ref{kernel-general}--\ref
{kernel-ext-L}) we see that 
\begin{eqnarray}
w_{4}=\left( n_{aa,cc}-n_{cc,aa}\right) /2.
\end{eqnarray}
In what follows we first demonstrate the presence of the first order phase
transition in the model of Eq. (\ref{D2}) and then provide a microscopical
derivation of $w_{4.}$

The critical magnetic field of the 90$^{\circ }$ reorientation $B_{reo}$
depends only on the ratio $r=-\lambda ^{2}w_{6}/w_{4}.$ We determined this
dependence numerically using standard computational methods. At first, for a
fixed $B$ the equilibrium form of VL unit cell was obtained by minimization
of Eq.(\ref{Eff-ener-pert}). Then, the zero of the perturbation energy Eq. (%
\ref{Eff-ener-pert}) was found. As usual \cite{ext-L} during the numerical
calculations the cutoff factor $\exp \left( -\xi ^{2}{\bf \tilde{g}}%
^{2}\right) $\ was introduced inside the above sums in order to properly
account for the failure of the London approach in the vortex core. The
calculated critical field is presented on Fig. 1 (we used $d=0.05$ and $%
n=0.015$ typical for ${\sl LuNi}_{2}{\sl B}_{2}{\sl C}$). We see that within
the approximation of Eq. (\ref{D2}) the $90^{\circ }$ reorientation cannot
happen at very low magnetic fields. For ${\sl LuNi}_{2}{\sl B}_{2}{\sl C}$%
{\sl \ } with $\lambda \approx 710\,$\AA , the field unit $\Phi _{0}/(2\pi
\lambda )^{2}$ is about $100\,G$. From the experimentally observed
transition field ${\sl B}_{reo}=2.95\,kOe$ \cite{90reo-neut} we estimate the
relative strength of sixth and fourth order terms in $D_{2}$ (see Eq. (\ref
{D2})) as $r=0.036.$


To obtain $w_{4}$ we start by discussing a general pairing model which
includes anisotropies in both the dispersion relation of electrons and the
singlet pairing interaction 
\begin{eqnarray}
H[\psi ] &=&\int_{x}\psi _{\alpha }^{\dagger }\left[ \varepsilon \left(
-i\nabla \right) -\mu \right] \psi _{\alpha }+V,  \label{Hamiltonian} \\
V &=&-\frac{\lambda }{4}\int_{x}\psi _{\alpha }^{\dagger }\left[ 1+\delta
\left( i\nabla \right) \right] \psi _{-\alpha }^{\dagger }\psi _{-\alpha }%
\left[ 1+\delta \left( -i\nabla \right) \right] \psi _{\alpha },  \nonumber
\end{eqnarray}
where the summation over spin indices $\alpha =\uparrow ,\downarrow $ is
assumed. Here $\psi (x)$ and $\psi ^{\dagger }(x)$ are the electron
destruction and creation operators, $\mu $ is chemical potential and $%
\lambda $ is a positive constant factorized from the pairing interaction for
convenience. Dispersion relation $\varepsilon ({\bf k})$ and pairing
interaction, of which $\delta ({\bf k})$ is a part, are usually defined in $%
{\bf k}$-space. To treat magnetic field effects it is advantageous to define
them in coordinate space. According to the rules of quantum mechanics, in
the above functions of ${\bf k}$ we perform the replacements ${\bf k}%
\rightarrow -i\nabla $ or ${\bf k}\rightarrow $ $i\nabla $ depending on
whether derivatives act on $\psi $ or $\psi ^{\dagger }$. Then the standard
minimal substitution $-i{\bf \nabla }\rightarrow {\bf \Pi \equiv }-i{\bf %
\nabla }-{\bf A}$ can be accomplished. This procedure, however, is not
unique because the components of ${\bf \Pi }$ do not commute with each
other. Therefore, $\varepsilon ({\bf k})$ and $\delta ({\bf k})$ are
presented by their Taylor expansions and in those terms which contain mixed
derivatives the symmetrization in $\Pi _{i}$ is used.

The kernel $K_{ij}({\bf q})$ from Eq. (\ref{kernel-general}) is obtained by
treating the effect of slowly varying magnetic field in terms of the linear
response. The change in the Hamiltonian due to the presence of magnetic
field $H_{1}[\psi ,{\bf A}]\equiv H[\psi ,{\bf A}]-H[\psi ,0]$ is taken into
account perturbatively. The result reads (see, for example, Ref. %
\onlinecite{Fetter}): 
\begin{eqnarray}
K_{ij}({\bf x-y})=\left\langle \frac{\partial ^{2}H_{1}}{\partial A_{i}({\bf %
x})\partial A_{j}({\bf y})}-\frac{\partial H_{1}}{\partial A_{i}({\bf x})}%
\frac{\partial H_{1}}{\partial A_{j}({\bf y})}\right\rangle ,
\label{kernel-micro}
\end{eqnarray}
where angular brackets denote the statistical average with unperturbed
density operator $\exp \left( -H[\psi ,0]/T\right) .$ Thus, we have to
expand the functional $H_{1}$ up to the terms quadratic in ${\bf A.}$
Because our aim is to calculate $w_{4}$ we need only the coefficients of 
this expansion for $A_{z} $ and $\partial A_{z}/\partial x$.

In its full generally the problem of Eq. (\ref{Hamiltonian}) in magnetic
field is quite intractable and below we consider two particular cases which
help us to estimate quantitatively the magnitude of different contributions
to $w_{4}$: i) isotropic superconducting interaction, $\delta ({\bf k})=0,$
and arbitrary dispersion relation $\varepsilon ({\bf k})$; ii) an example of
weakly ${\bf k}$ dependent superconducting electronic interaction, $\delta (%
{\bf k})=\delta _{0}k_{z}^{2},$ and isotropic dispersion of the standard
form $\varepsilon ({\bf k})={\bf k}^{2}/(2m)$. For simplicity in both cases
a clean system was investigated.

In the case i) we obtain 
\begin{eqnarray}
H_{1} &=&-\int_{{\bf x}}\psi _{\alpha }^{\dagger }\left[ A_{z}\varepsilon
_{,z}-i\left( \partial _{x}A_{z}\right) \frac{\varepsilon _{,zx}}{2}-\left(
\partial _{x}^{2}A_{z}\right) \frac{\varepsilon _{,zx^{2}}}{6}\right] \psi
_{\alpha }  \nonumber \\
&&+\frac{1}{2}\int_{{\bf x}}\psi _{\alpha }^{\dagger }\left[
A_{z}^{2}\varepsilon _{,z^{2}}-iA_{z}\left( \partial _{x}A_{z}\right)
\varepsilon _{,z^{2}x}\right.  \nonumber \\
&&\left. -A_{z}\left( \partial _{x}^{2}A_{z}\right) \frac{\varepsilon
_{,z^{2}x^{2}}}{3}-\left( \partial _{x}A_{z}\right) ^{2}\frac{\varepsilon
_{,z^{2}x^{2}}}{4}\right] \psi _{\alpha },  \label{S1-x}
\end{eqnarray}
where $\varepsilon _{,zx}$ means the second derivative of $\varepsilon (-i%
{\bf \nabla )}$ with respect to $z$ and $x$ components of the argument, and
so on. The final results reads 
\begin{eqnarray}
w_{4} &=&-\frac{1}{V}\sum_{{\bf k}}\left[ \frac{2}{3}R\,\varepsilon
_{,z}\varepsilon _{,zx^{2}}+\frac{\partial R}{\partial \varepsilon }%
\varepsilon _{,z}^{2}\varepsilon _{,x^{2}}-\left( x\leftrightarrow z\right) %
\right] , \\
R &=&\frac{4}{T}\cosh ^{-2}\left[ \frac{E({\bf k})}{2T}\right] ,\,E({\bf k})=%
\sqrt{(\varepsilon ({\bf k})-\mu )^{2}+\Delta ^{2}}.  \nonumber
\end{eqnarray}
At zero temperature $R$ approaches zero exponentially and $w_{4}$ vanishes.
As temperature increases, $w_{4}$ increases monotonically and reaches its
maximal value at $T=T_{c}$ where it smoothly joins the corresponding
component of $q$-dependent magnetic susceptibility tensor of the normal
metal. For estimation we considered a simple dispersion relation $%
\varepsilon ({\bf k})=\frac{{\bf 1}}{2m}{\bf k}^{2}+\frac{\tilde{\alpha}}{4}%
k_{z}^{4}$ and assumed deviations from spherical Fermi surface to be small: $%
\alpha \equiv \tilde{\alpha}m^{2}\mu \ll 1.$ Expanding in $\alpha $ we
obtain at $T=T_{c}$ that 
\begin{equation}
w_{4}^{FS}=2\alpha \Phi _{0}^{2}\sqrt{\hbar ^{2}\mu /2m}.
\end{equation}
where $\Phi _{0}=2e/hc.$ This quantity is very small. Indeed, comparing it
with the components of $n_{ij,kl}$ producing contributions to Eq.(\ref
{Ext-Lond-Energy}) we see that $w_{4}^{FS}/n_{xxxx}\sim \alpha \left( \Delta
/\mu \right) ^{2}.$ Therefore in order to find an origin of 90$^{\circ }$
reorientation one has to look elsewhere.

The obvious possibility is to relax the assumption of the isotropic gap and
turn to the case (ii). We calculated averages in Eq.(\ref{kernel-micro})
using the $1/N$ expansion\cite{large-N} rather than the BCS approximation.
The Hamiltonian Eq.(\ref{Hamiltonian}) becomes $\psi _{\alpha }^{a\dagger }%
\left[ \varepsilon \left( -i\nabla \right) -\mu \right] \psi _{\alpha }^{a}-%
\frac{\lambda }{4N}\psi _{\alpha }^{a\dagger }\left[ 1+\delta \left( i\nabla
\right) \right] \psi _{-\alpha }^{a\dagger }\psi _{-\alpha }^{b}\left[
1+\delta \left( -i\nabla \right) \right] \psi _{\alpha }^{b}$ where $N$ is
number of (real or auxiliary) copies of the Fermi surface enumerated by $a,$ 
$b$. The corresponding perturbation Hamiltonian found by the minimal
substitution is 
\begin{eqnarray}
H_{1} &=&i\int_{{\bf x}}A_{z}\left[ \frac{1}{2m}\psi _{\alpha }^{a\dagger
}\partial _{z}\psi _{\alpha }^{a}+\frac{\lambda \delta _{0}}{4N}%
S_{\downarrow \uparrow }^{\dagger }U_{\downarrow \uparrow }-cc\right] , \\
S_{\alpha \beta } &\equiv &\psi _{\alpha }^{a}\partial _{z}\psi _{\beta
}^{a}+\left( \partial _{z}\psi _{\alpha }^{a}\right) \psi _{\beta }^{a}, 
\nonumber \\
U_{\alpha \beta } &\equiv &\psi _{\alpha }^{a}\psi _{\beta }^{a}+\frac{%
\delta _{0}}{2}\left[ \psi _{\alpha }^{a}\partial _{z}^{2}\psi _{\beta
}^{a}+\left( \partial _{z}^{2}\psi _{\alpha }^{a}\right) \psi _{\beta }^{a}%
\right] .  \nonumber
\end{eqnarray}
Here the terms proportional to $A_{z}^{2}$ are omitted since they are local
and cannot contribute to derivatives of $K_{zz}(q)$ with respect to $q_{x}$
required to obtain $w_{4}$. For simpler situations like the case (i) the
leading order in $1/N$ expansion, with $N$ set to $1$, simply coincides with
the BCS approximation. The reason to resort to the $1/N$ expansion is
twofold. Firstly, the BCS expression for $w_{4}$ contains diagrams up to
three loops (see Fig. 2c) which are very complicated. Secondly, unlike BCS,
this nonperturbative scheme is systematically improvable. The last property
is important when questions of principle are concerned. After observing that
the order $1/N$ contributions, Fig. 2a, all vanish due to $%
k\Longleftrightarrow -k$ asymmetry, we calculated the leading $1/N^{2}$
contributions to the magnetic kernel, Fig. 2b. At $T=0$ to leading order in $%
\delta _{0}$ (further reducing number of integrals) the result reads 
\begin{equation}
w_{4}^{gap}=-\frac{\delta }{N^{2}}\frac{8\pi }{105}\left( \frac{\mu }{\Delta 
}\right) ^{2}\Phi _{0}^{2}\sqrt{\hbar ^{2}\mu /2m}
\end{equation}
where $\delta \equiv \delta _{0}m\mu $ is dimensionless gap anisotropy.
Therefore in physical case of interest $N=1$ we obtain $w_{4}^{gap}/n_{xxxx}%
\sim \delta $ that is not necessary very small. This value is to be compared
with $w_{4}^{FS}$ originated from Fermi surface anisotropy which has huge
suppression factor $(\mu /\Delta )^{2}.$ A noticeable angular dependence of
the gap was indeed observed in the most recent Raman scattering experiments
on $Y$ and $Lu$ borocarbides \cite{SILee}.


To conclude, we found that the extended London model is incapable of
explaining 90$^{\circ }$ reorientation in VL for $H||a$ because it produces
an effective fourfold symmetry of the free energy of VL. This symmetry
becomes explicit after a rescaling transformation. We showed that in general
case one should include into the extended London model correction terms for
which $n_{ij,kl}\neq n_{kl,ij}$ (see Eq. (\ref{kernel-ext-L})). As a result,
the true twofold symmetry of the system in magnetic field $H||a$ is restored
and 90$^{\circ }$ reorientation can be explained naturally. We demonstrated
the two mechanisms that generate the correction terms: anisotropy of the
Fermi surface and anisotropy of the superconducting gap, and showed that
only the contribution of the latter one can lead to observable consequences. 
The investigation of vortex
matter became recently a very sensitive tool to probe microscopic properties
of the superconductors. In this paper we employed it to infer qualitative
and even quantitative information about pairing interaction by calculating
nonlocal corrections to linear response.

Note that inclusion of the correction terms will not change any conclusions
of the extended London model for $H||c.$ On the other hand, nonzero ''two
fold symmetric'' correction will lead to smearing, or even disappearance, of
lock-in transition\cite{PTI} in VL for $H||a$ . Most probably it will be not
possible to check this prediction in the same samples of
${\sl LuNi}_{2}{\sl B}_{2}{\sl C}$ in which 90$^{\circ }$ reorientation was 
observed, because in this case the experimentally found opening angle of
the unit cell of rhombic VL\cite{90reo-neut} indicates that the critical field
of lock-in transition is far above $H_{c2}$.


We are grateful to V.G. Kogan for bringing this problem to our attention and
numerous illuminating discussions, to M.R. Eskildsen for discussions and
correspondence, and also to R. Joynt and Sung-Ik Lee for valuable comments.
The work is supported by NSC of Republic of China, grant $\#$%
89-2112-M-009-039.




\begin{figure}[tph]
\epsfig{figure=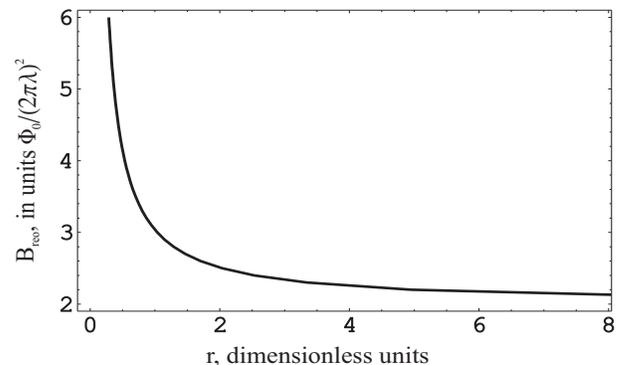,height=4.8cm,width=8cm}
\caption{Critical field of 90$^{\circ }$ VL reorientation as a function of
parameter $r=-\protect\lambda ^{2}w_{6}/w_{4}$ (see Eq. (10))}
\end{figure}

\begin{figure}[tph]
\epsfig{figure=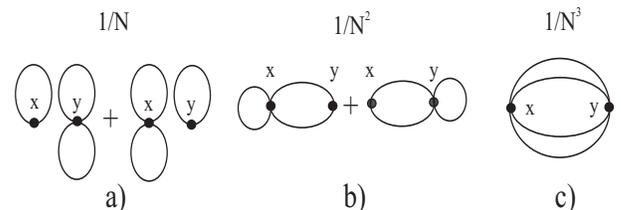,height=2.8cm,width=8cm}
\caption{
a) Diagrams of the order  $1/N$, which vanish in our model.
b) Diadrams of the order $1/N^2$ contributing to $w_4$. 
c) An example of complicated diagrams, which are of the order 
$1/N^3$ and were neglected. }
\end{figure}

\end{document}